\documentstyle[floats,prl,aps,epsf]{revtex}
\draft 
\begin{document}

%
%

\twocolumn[\hsize\textwidth\columnwidth\hsize\csname
@twocolumnfalse\endcsname
%

\title{Boosted three-dimensional black-hole evolutions with
	singularity excision}
\author{The Binary Black Hole Grand Challenge Alliance:} 
\author{G.~B.~Cook$^{\rm a}$, M.~F.~Huq$^{\rm b}$,
        S.~A.~Klasky$^{\rm c}$, M.~A.~Scheel$^{\rm a}$, \\
        A.~M.~Abrahams$^{\rm d,e}$, 
        A.~Anderson$^{\rm f}$, P.~Anninos$^{\rm d}$, 
        T.~W.~Baumgarte$^{\rm d}$, N.~T.~Bishop$^{\rm g}$,
        S.~R.~Brandt$^{\rm d}$, J.~C.~Browne$^{\rm b}$, 
        K.~Camarda$^{\rm h}$, M.~W.~Choptuik$^{\rm b}$, 
        C.~R.~Evans$^{\rm f}$, 
        L.~S.~Finn$^{\rm i}$, G.~C.~Fox$^{\rm c}$, 
        R.~G\'omez$^{\rm j}$, T.~Haupt$^{\rm c}$, 
        L.~E.~Kidder$^{\rm i}$, 
        P.~Laguna$^{\rm h}$, 
        W.~Landry$^{\rm a}$, L.~Lehner$^{\rm j}$, 
        J.~Lenaghan$^{\rm f}$, R.~L.~Marsa$^{\rm b}$,
        J.~Masso$^{\rm d}$, R.~A.~Matzner$^{\rm b}$, 
        S.~Mitra$^{\rm b}$, P.~Papadopoulos$^{\rm h}$, 
        M.~Parashar$^{\rm b}$, L.~Rezzolla$^{\rm d}$, 
        M.~E.~Rupright$^{\rm f}$, F.~Saied$^{\rm d}$, 
        P.~E.~Saylor$^{\rm d}$, 
        E.~Seidel$^{\rm d}$, S.~L.~Shapiro$^{\rm d}$, 
        D.~Shoemaker$^{\rm b}$, L.~Smarr$^{\rm d}$, 
        W.~M.~Suen $^{\rm k}$,
        B.~Szil\'agyi$^{\rm j}$, S.~A.~Teukolsky$^{\rm a}$, 
        M.~H.~P.~M.~van Putten$^{\rm a}$, P.~Walker$^{\rm d}$, 
        J.~Winicour$^{\rm j}$, J.~W.~York Jr$^{\rm f}$.}

\address{$^{\rm a}$Cornell University, Ithaca, New York 14853}
\address{$^{\rm b}$The University of Texas at Austin,  
                   Austin, Texas 78712}
\address{$^{\rm c}$Syracuse University, Syracuse, New York 13244-4100}
\address{$^{\rm d}$University of Illinois at
                   Urbana-Champaign, Urbana, Illinois 61801}
\address{$^{\rm e}$J.~P. Morgan, 60 Wall St., New York, New York 10260}
\address{$^{\rm f}$University of North Carolina, Chapel Hill, 
North Carolina 27599}
\address{$^{\rm g}$University of South Africa, P.O. Box 392, 
                   Pretoria 0001, South Africa}
\address{$^{\rm h}$Penn State University, University Park, Pennsylvania 16802}
\address{$^{\rm i}$Northwestern University, Evanston, Illinois 60208}
\address{$^{\rm j}$University of Pittsburgh, Pittsburgh, Pennsylvania 15260}
\address{$^{\rm k}$Washington University, St. Louis, Missouri 63130}
\maketitle
\begin{abstract}
Binary black hole interactions provide potentially the strongest
source of gravitational radiation for detectors currently under
development. We present some results from the Binary Black Hole Grand 
Challenge Alliance three-dimensional Cauchy evolution module. These 
constitute essential steps towards modeling such
interactions and predicting gravitational radiation waveforms. We
report on single black hole evolutions and the first successful
demonstration of a black hole moving freely through a three-dimensional
computational grid via a Cauchy evolution: a hole moving 
$\sim 6M$ at $0.1c$ during a total evolution of duration $\sim 60M$.
\end{abstract}


\vskip2pc]

The accurate computational modeling of black-hole interactions is
essential to the confident detection of astrophysical gravitational
radiation by future space-based detectors such as LISA and by the
LIGO/VIRGO/GEO complex of ground-based detectors currently under
construction.  The sensitivity of these detectors will be
significantly enhanced if accurate computer simulations of black-hole
mergers can produce predictions of radiation
waveforms~\cite{flanagan.hughes}.  The Binary Black Hole Grand
Challenge Alliance~\cite{bbh} was funded in September 1993 to develop
the computational infrastructure necessary accurately to simulate the
coalescence of black-hole binaries.  The primary objective of the
resulting code will be the production of waveforms from binary black
hole mergers. In this Letter we report on an important step towards
achieving such simulations.

A key difficulty in evolving black-hole spacetimes is handling the
curvature singularity contained within each hole.  The only viable
means of accomplishing this over time scales required for binary
coalescence appears to be black-hole excision: exclude all or part of
the black-hole interior (and the singularity) from the computational
domain and evolve only the exterior region\cite{thornburg,unruh}.  This is 
possible because, by definition, the region inside the hole cannot causally 
affect the exterior evolution. Black-hole excision has been implemented
successfully in spherical symmetry by Seidel and
Suen~\cite{seidel.suen}, Scheel {\it et
al.}~\cite{scheeletal1,scheeletal2}, Marsa and
Choptuik~\cite{choptuik.marsa}, Anninos {\it et al.}\cite{anninos}, and 
Gomez {\it et al.}
\cite{gomez}; and in three dimensions (3D) by Daues~\cite{daues.phd} and by
Gomez {\it et al.}\cite{gomez2}.

We are developing a general algorithm to perform a Cauchy evolution of
Einstein's equations on a domain containing black holes with excised
interiors.  Prior simulations of black-hole binaries have used
coordinates in which the holes remain at fixed coordinate locations;
for complicated motions, this will lead to undesirably contorted
coordinates. In contrast, our more general approach allows black holes
with excised interiors to move freely through the computational
grid. Achieving this goal requires the synthesis of two key
ingredients: a numerical scheme capable of stably evolving Einstein's
equations on a domain containing moving boundaries (excised regions),
and a set of gauge conditions that ensure that coordinates remain
well-behaved as black holes move through the grid. In this Letter we
present a crucial step towards achieving a general black-hole
evolution scheme: the first successful Cauchy computation of a single
black hole freely moving through a 3D numerical grid. A different 
approach within the Alliance that uses a characteristic formulation
\cite{gomez2} has recently shown 3D black-hole evolutions to $1400M$ and 
success in moving a black hole less than one diameter.

Black-hole excision is based on the fundamental idea that the event
horizon of a black hole is a natural causal boundary.  Unfortunately,
an event horizon cannot be located without knowing the full future
evolution of the spacetime. However, a related structure known as an
apparent horizon can usually be located on a spacelike time slice
using only the information on that hypersurface.  An apparent horizon
is a topologically spherical spacelike two-surface on which the
expansion of the congruence of outgoing null rays orthogonal to the
surface vanishes.  If the characteristic curves of all fields being
evolved lie on or within the light cone, then the apparent horizon can
be used as the {\em inner boundary} for a Cauchy evolution.  The
causal nature of this boundary implies that no explicit boundary
condition need be imposed on the evolved quantities.

The gauge freedom (coordinate freedom) of general relativity allows
considerable latitude in choosing how a computational solution evolves
in time.  Although gauge considerations cannot influence physics, they
do determine how the coordinates and computational grid points used to
describe the solution relate to physical locations in spacetime.  A
poor choice of spatial or temporal gauge can lead to coordinate
pathologies that ruin a numerical simulation. For example, the proper
distance between two adjacent computational grid points might approach
zero or grow without bound.  It is not fully understood what
constitutes a {\em good} gauge choice.

An attractive choice\cite{allenetal96,choptuik.marsa} for describing a
single black hole is the ingoing Kerr-Schild form\cite{mtw}
of the Kerr metric:
\begin{eqnarray}
        ds^{2} = g_{\mu \nu} dx^{\mu} dx^{\nu} = 
        \left( \eta_{\mu \nu} + 2H l_{\mu} l_{\nu} \right) 
	dx^{\mu} dx^{\nu},
\label{kerrschild.metric}
\end{eqnarray}
where $\mu, \nu$ run from $0$ to $3$, $\eta_{\mu \nu} =
\mbox{diag}(-1,1,1,1)$, $l^{\mu}$ is a null four-vector, and
$H(x^{\alpha})$ is a scalar function.  In this gauge, the coordinates
are closely related to the null structure of the
spacetime. Furthermore, the solution is time independent (or has a
trivial time-dependence for a moving black hole), the spacelike
hypersurfaces extend smoothly through the horizon, and gradients near
the horizon are smaller than in several other coordinate choices.

Equation (\ref{kerrschild.metric}) is form-invariant under Lorentz
transformations, so it can be used to represent either a
non-translating or a boosted Kerr black hole. The 3+1 decomposition
of the spacetime metric leads to a metric $g_{ij}$ and extrinsic
curvature $K_{ij}$ that we use as initial data and for comparisons at
later times\cite{huq}.  The lapse function,
\begin{eqnarray}
       \alpha = 1 / \sqrt{1 + 2 H l_t^2},
\label{kerrschild.lapse}
\end{eqnarray}
and shift vector,
\begin{eqnarray}
       \beta_i = 2 H l_t l_i,
\label{kerrschild.shift}
\end{eqnarray}
are analytic functions of space and time that we impose as gauge
conditions.  For the time-independent Schwarzschild spacetime, $H=M/r$
($M$ is the mass), $l_{\mu} = (1,x_i/r)$, and the apparent horizon
coincides with the event horizon.  In fact, the apparent and event
horizons coincide in the boosted case as well.  

Schemes for excising the interior of a black hole from the
computational grid typically require a superluminal shift vector 
in some region of the computational domain, and must cope with
the lack of an explicit boundary condition on the excision boundary.
The Alliance 
Cauchy evolution module implements an evolution scheme that is designed to 
provide a stable evolution for any choice of shift vector.
A typical evolution
equation has the form
\begin{equation}
\label{evolution_generic}
\left(\partial_t - \pounds_\beta\right)T = \ldots,
\end{equation}
where $\pounds_\beta$ is the Lie derivative along the shift vector and
$T$ is an arbitrary spatial tensor.  
We can rewrite this as
\begin{equation}
\label{evolution_normal}
\partial_0 T -\left(\pounds_\beta - \beta^i\partial_i\right)T = \ldots,
\end{equation}
where $\partial_0\equiv\partial_t - \beta^i\partial_i$ is a time
derivative in the direction normal to the spatial hypersurface and
which, by definition, is centered in the light cone.  This scheme
removes from $\pounds_\beta$ the advective term that potentially leads
to evolution along a non-timelike direction. This guarantees that a
numerical evolution will be stable against instabilities produced by
superluminal shifts.  Note that the tensor components being evolved
remain in the coordinate basis $\partial/\partial x^i$ that is Lie
dragged along the $t^\mu$ direction, but the computational grid points at
which these components are defined do not remain at constant values of
$x^i$. Instead, the grid points remain at constant values of the
spatial coordinates $\tilde{x}^i$ that are Lie dragged along the unit
normal to the spatial hypersurface.  The evolved quantities are
determined at the desired spatial coordinate locations $x^i$ by
interpolation.  This interpolation requires that the $\tilde{x}^i$
coordinate values be evolved along the $t^\mu$ direction for each
point where the evolved quantities are to be evaluated on a given time
slice.  
It also obviates the need for a boundary condition at the excision surface.
The algorithm can easily accommodate black holes that move
through the grid.  In this case, unused grid points that had been in
the interior of a hole can move into the exterior.  These new points
will be filled by the interpolation phase and will always be filled
from data that is in the future domain of dependence of the previous
time slice. 
Similar schemes were developed by Seidel and
Suen~\cite{seidel.suen}, Alcubierre and Schutz~\cite{alcubierre94},
and Daues~\cite{daues.phd}.  Additional details on how this evolution
scheme is implemented can be found in Scheel {\it et
al.}~\cite{scheeletal2} and in a forthcoming paper\cite{cook97} that will 
describe
the details of the Alliance Cauchy evolution module.  The key point is
that by splitting the evolution along $t^\mu$ into an evolution along
the unit normal followed by an interpolation, the algorithm guarantees
that the numerical evolution step is always taken in the center of the
physical light cone.

The Cauchy evolution module developed by the Alliance is based on the
standard (3+1), or ADM, decomposition of Einstein's field equations
\cite{adm62,york79}.  The module uses Cartesian coordinates and
evolves the three-metric $g_{ij}$, and extrinsic curvature $K_{ij}$, on
a 3D rectangular grid. The evolution equations are
solved using an iterative Crank-Nicholson differencing scheme. 
The constraint equations are not imposed on the solution but are utilized
as a diagnostic.

We present results from two sets of black-hole experiments: evolutions
of non-translating and of boosted Schwarzschild black holes. In both
sets of experiments we work with a 3D $n\times n\times m$
computational grid, and we impose {\em Dirichlet} boundary conditions
on the grid faces throughout the evolutions.  For the non-translating
cases, this means freezing the outer boundaries to initial data.  For
the boosted case, this means resetting the exact solution at the faces
for each time step.  At each time during the course of our evolution
we impose the exact gauge conditions (\ref{kerrschild.lapse}) and
(\ref{kerrschild.shift}) as a function of four spacetime coordinates.
We allow for a ``buffer region'' of $p$ grid zones between the apparent
horizon (where the inner boundary might have been placed) and the 
chosen inner boundary (excised region).

\begin{figure}
\epsfxsize=7.5cm
\begin{center}
\leavevmode
\epsffile{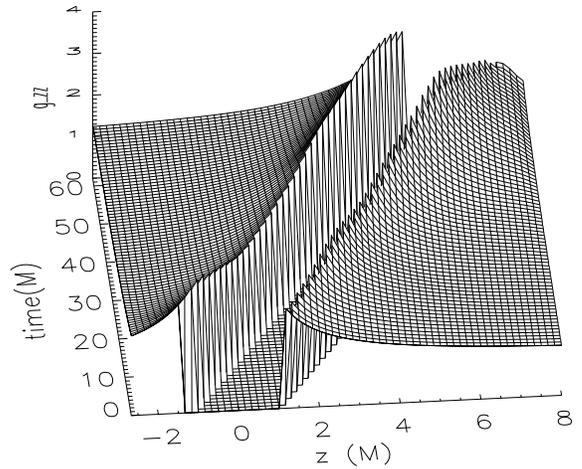}
\end{center}
\caption{Metric component $g_{zz}$ along the $z$-axis as a function of
time. The flat region that moves diagonally to the right represents the
excised region (inside the black hole).  Note that points at
the trailing edge (left side) are smoothly updated as the hole moves towards
positive $z$. Coordinate effects are seen to appear near the inner boundary.}
\label{boosted.fig}
\end{figure}

\begin{figure}
\epsfxsize=7.5cm
\begin{center}
\leavevmode
\epsffile{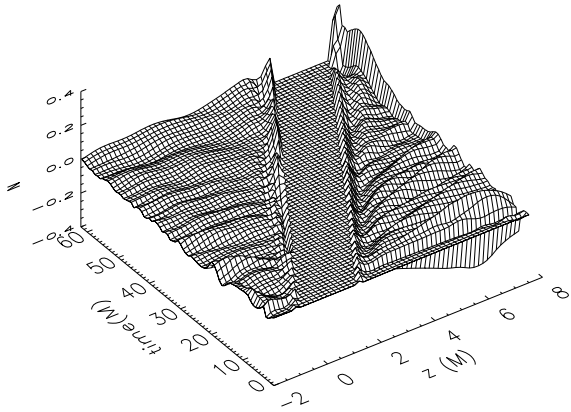}
\end{center}
\caption{Normalized Hamiltonian constraint, $N = (R + K^2 - 
K^{ij}K_{ij})/(|R| + |K^2| + |K^{ij}K_{ij}|)$, along the $z$-axis as
a function of time. The flat region shows the location of the
excised part of the domain within the hole.}
\label{ham.fig}
\end{figure}

For the case of a boosted black hole we discuss two runs; one with
$p=3$ buffer zones and the other with $p=5$ buffer zones between the
horizon and the inner boundary. Both runs use $n=33$ and $m=65$ with a
domain of $-(8/3)M$ to $8M$ in the $z$-direction and a domain of $-(8/3)M$
to $(8/3)M$ in the $x$ and $y$ directions.  The black hole is initially
located at the origin and has a velocity $v=0.1c$ in the $z$
direction.  The evolution was carried out with a Courant factor of
$1/4$.  Figure~\ref{boosted.fig} shows a spacetime picture of $g_{zz}$
along the $z$-axis vs time for the $p=5$ run.  In this case we evolved
the black hole for $61M$, moving three hole radii. The evolution
terminated when the leading edge of the black hole was within five
gridpoints of the $z=8M$ face. At this point the differencing
algorithms failed because of an insufficient number of points between
the inner and outer boundaries. In the $p=3$ run the black hole
evolved for $54M$ with coordinate stretching occurring at the trailing
edge.  We find that the addition of buffer zones enhanced the run and
reduced coordinate stretching for the duration of
the $p=5$ run.  This is consistent with the behavior of the
non-translating runs described below.  As the hole moved in each of
the evolutions, many coordinate points emerged stably and smoothly
from the excised region into the computational domain.
Figure~\ref{ham.fig} shows a spacetime picture of the normalized
Hamiltonian constraint diagnostic along the $z$-axis for the $p=5$
run.  Note that the Hamiltonian constraint is well behaved in the
regions where grid points have emerged from the black hole.

For the case of a non-translating black hole, we chose $n=m=49$. The
outer boundaries are placed at $-4M$ and $4M$ in each direction.  With
a Courant factor of $1/4$ we can evolve a single black hole to $95M$.
This is an encouraging achievement comparable to the longest 3D
single black hole Cauchy simulation ($>100M$ ; Daues \cite{daues.phd}).

As in the moving hole case, the length of the evolutions is dependent
on the placement of the inner boundary.  We found that by setting the
number of buffer zones to $p=0$, $1$, $5$, and $9$, we can run a
non-moving hole to a maximum time of $t_{max} = 16M$, $20M$, $95M$,
and $82M$.  In all cases, the evolution terminates because the
iterative Crank-Nicholson evolution scheme fails to converge to a
solution.  While the evolved solutions all deviated from the analytic
solution, the behavior in each case was somewhat different and a
definitive explanation of why each evolution fails requires further
testing.  The most likely causes of the late time problems are either
coordinate effects and/or numerical instabilities due to choices of
finite difference operators.  One such coordinate effect can come from
fixing the gauge via the analytic functions (\ref{kerrschild.lapse})
and (\ref{kerrschild.shift}).  As numerical evolutions progress, the
evolved data drift from the exact solution.  When this happens, the
choice of lapse and shift being used will no longer enforce the
desired, underlying Kerr-Schild gauge condition.  In this case,
effects such as grid stretching can occur and the coordinate location
of the apparent horizon will no longer coincide with the analytic
solution.  For $p=0$ or $1$, the evolution progresses rather smoothly
and the computed location of the apparent horizon gradually moves
inward.  Eventually the apparent horizon passes within the
computational inner boundary, so that the inner boundary becomes
timelike and therefore unsuitable as a boundary for black-hole
excision.  The evolutions terminate soon after this occurs.  For $p=5$
or $9$, the evolution is more complicated.  The relatively smooth
growth in error with the apparent horizon moving inward no longer
occurs (or occurs on a much longer time scale).  In these cases, the
dominant errors appear to be noise introduced at the Dirichlet outer
boundaries.  These errors propagate across the computational grid and
appear to be amplified in some way.  It is possible that the analytic
gauge conditions are causing this amplification; some unusual coupling
with the inner boundary may also be responsible.  Eventually, the
geometry near the inner boundary becomes quite non-spherical.  While
we are no longer certain of the location of the apparent horizon at
this point, we believe that the inner boundary again becomes timelike
prior to the evolution ending.

The fact that the evolutions depend strongly on the placement of the
inner boundary agrees with prior 3D work by Daues \cite{daues.phd} but
requires further study.  Using a larger number of buffer zones forces
the inner boundary deeper into the black-hole interior where an
increasingly larger propagation speed is required for information to
escape from the black hole.  Thus the strong dependence of the
evolutions on $p$ could indicate some non-physical gauge-dependent
quantities or numerical errors propagating faster than light.
Resolution of this question must await at least a consistent set of
convergence tests to understand the effect of $p$ as the
discretization is refined.  In spite of these concerns, the evolution
to $95M$ is an encouraging result and we believe that improved gauge
choices that utilize information from the evolution ({\it e.g.\/} the
computed apparent horizon location,
cf.\ Refs.~\cite{scheeletal2,daues.phd}) will allow for much longer
evolutions.

In the evolutions of non-translating holes, each case above results in an
unstable evolution.  We believe that this is most likely due to
coordinate effects.  This belief is supported by evolving a 
region adjacent to a non-translating black hole.  We consider a domain
extending from $-(3/2)M$ to $(3/2)M$ in all directions and displace the
center of the black hole along one axis by $2M$.  
By this design the horizon passes through the center of the computational
cube.
Using resolution $n=m=49$ and Dirichlet boundary conditions on all
boundaries, we find that the solution settles down to a steady state
because of numerical dissipation and
perhaps because wave disturbances fall within the horizon.
At low enough resolution, the errors in the analytic gauge conditions
and Dirichlet boundaries were too large to allow a steady-state
solution.  As a rule, higher resolution cannot force a fundamentally
unstable numerical scheme to be ``more stable''.  We thus find this
test to be highly suggestive that our interior evolution scheme is
fundamentally stable and our principal problems are with coordinate and
boundary effects.

All of the above results are very encouraging for the ultimate goal of
evolving multiple black hole spacetimes. In both the boosted and
non-translating evolutions, improvements to the coordinate conditions
are clearly necessary.  Non-analytical coordinate conditions need to
be developed for the boosted case and we are hopeful that it will be
possible to formulate a general Kerr-Schild-like gauge condition that
does not require the solution of elliptic equations.  Other
combinations of spatial and temporal gauge conditions may also work or
may even be necessary. There is still considerable work to be done on
the computational infrastructure required to support general binary
black hole simulations, but it is clear that one of the keys to
successful evolutions is understanding which gauge conditions are
appropriate for multiple black holes.

This work was supported by the NSF Binary Black Hole Grand Challenge
Grant Nos. NSF PHY/ASC 93--18152 (ARPA supplemented), NSF PHY
93--10083, and Metacenter Grant MCA94P015N.  Computations were
performed at NPAC (Syracuse University), at NCSA (University of
Illinois at Urbana-Champaign) and at the Pittsburgh Supercomputing
Center.

\end{document}